\documentclass[prl,aps,twocolumn,superscriptaddress,showpacs]{revtex4}
\usepackage{graphics,graphicx}

\begin{document}

\title{      Ferro-Orbitally Ordered Stripes
       in Systems with Alternating Orbital Order}

\author{     Piotr Wr\'obel}

\affiliation{Max Planck Institute for the Physics of Complex Systems,
             N\"othnitzer Stra{\ss}e 38, 01187 Dresden, Germany}
\affiliation{Institute for Low Temperature and Structure Research,
             P.O. Box 1410, 50-950 Wroc{\l}aw 2, Poland}

\author{     Andrzej M. Ole\'s}
\affiliation{Max-Planck-Institut f\"{u}r Festk\"{o}rperforschung,
             Heisenbergstrasse 1, 70569 Stuttgart, Germany}
\affiliation{Marian Smoluchowski Institute of Physics, Jagellonian University,
             Reymonta 4, 30-059 Krak\'ow, Poland}

\date{ 28 January 2010 }

\begin{abstract}
We establish a novel mechanism of stripe formation in doped systems
with alternating $t_{2g}$ orbital order --- the stripe takes the form
of a ferro-orbitally ordered domain wall separating domains with
staggered order and allowing for deconfined motion of holes along
the stripe. At a finite level of hole concentration this gives rise
to the stability of this solitonic type of stripes, while we show that
the phase change of the staggered order by $\pi$ plays a minor role in
orbitally ordered systems. These results shed new light on the physics
of doped materials in which orbital degeneracy is present.

{\it Published in: Phys. Rev. Lett.} {\bf 104}, 206401 (2010).
\end{abstract}

\pacs{71.10.Fd, 72.10.Di, 72.80.Ga, 75.60.Ch}

\maketitle

\begin{figure}[b!]
 \centering
\includegraphics[width=8.4cm]{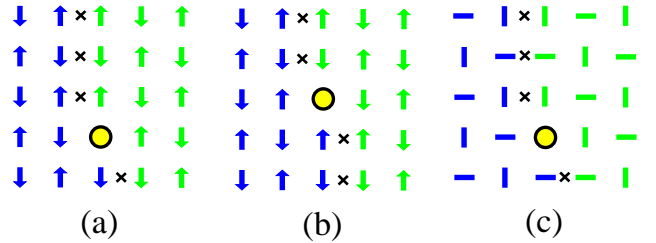}
\caption{(color online)
The mechanism of stripe formation in the spin Ising model:
a hole doped at the DW kink (a) moves together with the kink (b).
In contrast, a hole doped at the DW kink in the orbital $t_{2g}$
system (c) is confined to two sites.
Two domains with AF (AO) order are shown by arrows (boxes);
broken bonds are marked by $\times$.}
\label{Fig1}
\end{figure}

In the context of superconductivity in cuprates both experimental
and theoretical aspects of stripes have been the subjects of
intensive research \cite{strip}. At least in the low doping range,
these systems can be viewed as an antiferromagnetic (AF) phase
into which holes have been injected.
When a hole hops in an AF background, it interchanges position with
a single spin in each step and creates a ``string'' of flipped spins
along its path \cite{Bul68}, which accumulates energy cost and thus
confines the hole in the system with classical (Ising-type)
interaction. At finite doping spin and charge density modulations
(stripes) develop, which is a way to find compromise between
two opposite tendencies:
(i) to delocalize holes and gain hopping energy $\propto t$, and
(ii) let AF correlations to develop, which optimize
the superexchange energy $\propto J$.
It seems that due to the presence of quantum spin fluctuations stripes
in cuprates show bond order, i.e. take the form of ladders with
dominating singlet correlations on the rungs \cite{Wro06}. In addition,
stripe formation has been demonstrated to exist in a model with classical
AF exchange interaction \cite{Che00}, so called $t$-$J_z$ model,
outlined in Figs. \ref{Fig1}(a) and \ref{Fig1}(b). While the $t$-$J_z$
model is hard to realize in spin systems, we show in this Letter that
a related mechanism of stripe generation would work
in $t_{2g}$ orbital systems.

Some previous theoretical analyzes of stripe formation in
orbitally degenerate systems lead to the conclusion that lattice
distortions are essential for this kind of ordering in systems
with $e_g$ orbital degrees of freedom \cite{Hot01}. Stripes in
pnictides with active $t_{2g}$ orbitals were suggested only very
recently \cite{Jan09}. We will demonstrate that in systems with
$t_{2g}$ degeneracy and for large onsite Coulomb interaction $U$
merely the interplay between hopping and the orbital superexchange
interaction gives rise to the formation of ferro-orbitally (FO)
ordered stripes as domain walls (DWs) between regions with the
alternating orbital (AO) order. With the goal of analyzing this
phenomenon in detail we concentrate on the recently introduced
strong-coupling version of the multi-orbital Hubbard model for
spinless fermions \cite{Woh08} on the square lattice (when the
spins form a ferromagnetic order).
This model is applicable either to transition metal oxides with
active $t_{2g}$ orbitals (when the tetragonal crystal field splits
off the $xy$ orbital from the $\{yz,zx\}$ doublet filled by one
electron at each site,
as for instance in Sr$_2$VO$_4$ \cite{Mat05}), or to cold-atom
systems \cite{Jak05} with active $p$ orbitals \cite{Lu09}.

The strong-correlation limit of the model reads:
\begin{eqnarray}
\label{t2gmodel} {\cal H}_{t_{2g}} &=&{\cal P}\;\left( {\cal
H}_{t}+{\cal H}_{J} +{\cal H}_{\rm 3s}^{(l)} +{\cal H}_{\rm
3s}^{(d)}\right)\;{\cal P}\,,\\
\label{Ht} {\cal H}_{t}&=&-t \sum_{ i}
({b}^{\dagger}_{i}{b}^{}_{i+\bf{\hat{a}}} +
{a}^{\dagger}_{i}{a}^{}_{i+\bf{\hat{b}}}+\mbox{H.c.})\,,\\
\label{HJ} {\cal H}_{J}&=& \frac12 J \sum_{\langle ij\rangle }
\left(T^z_i T^z_j - \frac{1}{4}{n}_i{n}_j\right)\,,\\
\label{H3sl} {\cal H}_{\rm 3s}^{(l)} &=& -\tau \sum_{i}
({b}^\dag_{i-\bf{\hat{a}}}{n}^{}_{ia}
{b}^{}_{i+\bf{\hat{a}}}+\mbox{H.c.}) \nonumber\\
&& -\tau\sum_{i}({a}^\dag_{i-\bf{\hat{b}}}
{n}^{}_{ib}{a}^{}_{i+\bf{\hat{b}}}+\mbox{H.c.})\,,\\
\label{H3sd}
H_{\rm 3s}^{(d)} &=&-\tau \sum_{i} ({a}^\dag_{i\pm\bf{\hat{b}}}{a}^{}_i
{b}^\dag_i{b}^{}_{i\pm\bf{\hat{a}}}+\mbox{H.c.}) \nonumber\\
&&-\tau \sum_{i} ({a}^\dag_{i\mp\bf{\hat{b}}}{a}^{}_i
{b}^\dag_i{b}^{}_{i\pm\bf{\hat{a}}}+\mbox{H.c.})\,.
\end{eqnarray}
Here $a$ and $b$ refer to two $t_{2g}$ orbital flavors
\cite{Woh08}, $yz\equiv a$ and $zx\equiv b$, and the summations
are carried over $i\in ab$ sites in the $ab$ plane. The orbital
superexchange
$J=4t^2/U$ and the effective next
nearest neighbor hopping $\tau=t^2/U$ apply when $U\gg t$ \cite{noteu}.
The pseudospin operator is
$T^z_i=\frac{1}{2}\left({n}_{ia}-{n}_{ib}\right)$, while the projection
operator ${\cal P}$ removes from the Hilbert space states in which any
site is doubly occupied; for more details see Ref. \cite{Woh08}.

Single hole motion is in principle confined in the spin $t$-$J_z$ model
due to the potential well effect caused by the formation of strings.
An effective way to avoid the string effect is to form an antiphase DW
between two AF domains, consisting of two semilines separated by a
transversal kink, and to create a hole at one of two sites nearest to
the kink in that DW, see Figs. \ref{Fig1}(a)-(b). The hole may be
shifted along the wall without increasing the number of broken bonds.
As the energy gain of $-t$ is typically larger than the energy loss
$J_z/2$ due to a broken bond, at a certain level of hole doping the
energy decrease induced by free hopping of holes along the stripe will
compensate the increase of the magnetic energy caused by the creation
of the DW \cite{Che00}.

Despite the similarity between the spin $t$-$J_z$ model \cite{Che00}
and the $t_{2g}$ orbital model given by Eqs. (1-5), the mechanism
of stripe formation based on soliton-like motion of the kink-hole
complex is not applicable to orbitally degenerate systems. This
can be understood by analyzing the DW shown in Fig. \ref{Fig1}(c).
Boxes aligned along the $\hat\textbf{a}$ ($\hat\textbf{b}$)
direction represent $b$ ($a$) orbital flavors in two domains with
opposite phases of the AO order. Again, similarly to the Ising AF
state, we have created an antiphase DW with a kink, and we have
removed an electron from one of two sites nearest to the kink
center. The downward shift of the hole by one lattice spacing is
blocked by the $b$ orbital below it. This follows directly from
the form of the hopping (\ref{Ht}), which is one-dimensional (1D)
in ordered $ab$ planes \cite{Woh08}. The upward shift of the hole
by one lattice spacing is allowed, but after that move the hole
will be blocked again from above. Therefore, the hole and kink
motion are confined in the orbitally degenerate system with the
straight antiphase DW and a single kink in it [Fig.
\ref{Fig1}(c)]. Nevertheless, since the term (\ref{Ht}) brings the
biggest energy scale ($t\gg J$) we may expect that it will modify
the form of the ground state above a certain filling level. The
mechanism of hole deconfinement, however, is different as we show
below by a detailed stability analysis.

\begin{figure}[t!]
 \centering
 \includegraphics[width=7.8cm]{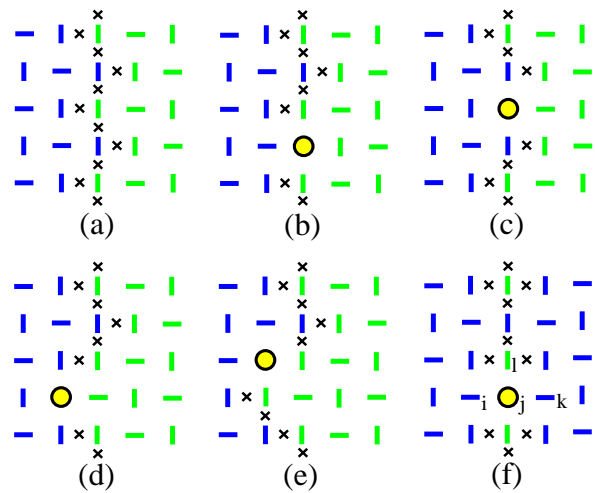}
\caption{(color online) Processes governing hole propagation in
the stripe formed as a DW between two domains of the AO order (a):
a doped hole may hop either along the DW (b)-(c) or sidewards (d)-(e);
(f) shows the PW of $a$ orbitals formed within a single AO domain;
broken bonds are marked by $\times$.}
\label{Fig2}
\end{figure}

The DW depicted in Fig. \ref{Fig1} is the most favorable one in terms
of the minimal number of broken bonds per one DW site. For an orbitally
degenerate system, however, the hole motion by hopping $t$ is allowed
along the chain only when orbitals reorient, similarly to the 1D
$e_g$ systems \cite{Dag04}. Such a FO ordered vertical chain of $a$
orbitals in Fig. \ref{Fig2}(a) provides a DW between AO domains, and
makes it possible to deconfine the hole motion along it. The price one
has to pay are two (not one) broken bonds per site. After replacing one
electron (orbital) in the chain by a hole [Fig. \ref{Fig2}(b)], the hole
can move due to ${\cal H}_t$ (\ref{Ht}) by one step [Fig. \ref{Fig2}(c)],
and two broken bonds are removed and two other ones are created ---
hence the total number of broken bonds remains unchanged. In this way
the hole motion occurs in both directions,
leaving behind the undisturbed zig-zag pattern of broken bonds.

The hole may also penetrate into the orbitally ordered domains,
as for instance when the term (\ref{Ht}) interchanges the hole with
$b$-orbital in Fig. \ref{Fig2}(b), leading to the state depicted in
Fig. \ref{Fig2}(d). Now the hole can hop further to a nearest neighbor
site either downwards or upwards, but when it enters ``deeper'' into
the AO domain, the number of broken bonds increases (now by one) and
the string effect \cite{Bul68} occurs [Fig. \ref{Fig2}(e)].
This mechanism efficiently confines the hole motion to the stripe DW.
The expected energy gain due to hole motion over a homogenous
AO state is realized by the parallel alignment of $a$-orbitals
in the chain, while the phase change by $\pi$ between two AO domains is
irrelevant to achieve that gain. A similar FO ordered chain deconfining
the hole motion and having the same energy cost can also be created by
reversing every second orbital along a vertical polaronic line within a
single domain of the AO order, see Fig. \ref{Fig2}(f) --- it generates
a polaronic wall (PW) by a mechanism similar to that which operates in
the 1D $e_g$ orbital model \cite{Dag04}. The main difference between
stripes depicted in Figs. \ref{Fig2}(b) and \ref{Fig2}(f) is that the
hole in the DW stripe may penetrate one domain from each stripe site,
while in the PW stripe it can enter both domains from every second site
[the hole in Fig. \ref{Fig2}(f) can move only from site $j$
(not from $l$) either to site $i$ or to site $k$].
The explicit analysis presented below will demonstrate that it is
energetically somewhat more favorable to create
the antiphase DW stripe.

\begin{figure}[t!]
 \centering
 \includegraphics[width=7.8cm]{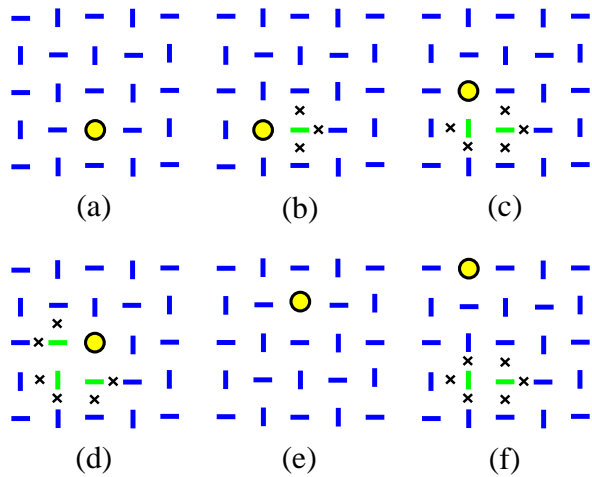}
\caption{(color online)
Artist's view of
hole propagation in the homogenous $|{\rm AO}\rangle$ state (a)
--- a hole may move due to $t$ (\ref{Ht}) and create broken bonds,
marked by $\times$, after one (b), two (c), or three (d) hops,
while $\tau$ (\ref{H3sl}) does not disturb the AO order (a) and (e),
but may also leave behind broken bonds (c) and (f).
}
\label{Fig3}
\end{figure}

In order to discuss stripe stability we must first analyze single hole
propagation in the homogenous AO phase by means of the same method
which will be used later
for stripes. We begin with a hole replacing $a$ orbital [Fig.
\ref{Fig3}(a)] --- the hopping term $\propto t$ can shift a hole
only horizontally to its neighboring sites which are occupied by
$b$ orbitals [Fig. \ref{Fig3}(b)]. The hole motion can be
continued and after the second and the third hop the number of
broken bonds increases, see Figs. \ref{Fig3}(c)-(d). In general,
the energy rise $\propto J$, growing with the number of broken
bonds, confines the hole motion (here initiated by the hop in the
direction $\hat\textbf{a}$) \cite{Woh08}. The number of defected
bonds can be easily evaluated explicitly for paths up to length of
seven steps, as we have done in our analysis. For longer paths
reasonable approximations can be developed. They follow
from the relation between the number of broken bonds and the number of
bends in the path, see Fig. \ref{Fig3}(d), and on the number of zigzags,
such as that arising if the hole in Fig. \ref{Fig3}(c) has moved left.

The action of the term (\ref{H3sd}) does not give rise to hole
deconfinement, as it moves a hole {\it generating several broken
bonds\/}, cf. Figs. \ref{Fig3}(a)-(c). In contrast, weak
hole deconfinement ($\tau\ll t$) occurs due to the term (\ref{H3sl})
which shifts a hole by two lattice spacings. This takes place without
bringing about any additional defects in the AO order and is allowed
provided that the opposite orbital (here $b$) occupies the intermediate
site [Figs. \ref{Fig3}(a), \ref{Fig3}(e)].

We are going now to cast the insights which have been outlined
above into the framework of the recursion method \cite{Sta94}
applied to the Green's function, and to determine the self-energy.
The starting point is the bare Green's function
$G_0(k_b,\omega)$ related with hole deconfined movement in the
$|{\rm AO}\rangle$ state mediated by the free propagation term
(\ref{H3sl}) along the $\hat\textbf{b}$ axis, with momentum $k_b$,
[Fig. \ref{Fig3}(e)] --- it is
$G_0^{-1}(k_b,\omega)\equiv\omega-2\tau\cos(2 k_b)-J$.
A reference
energy of the $|{\rm AO}\rangle$ state has been subtracted
in $G_0^{-1}(k_b,\omega)$, and the energy $J$ above arises from the
four bonds removed from the $|{\rm AO}\rangle$ state by adding a hole.
The energy dispersion $\propto\tau$ in $G_0(k_b,\omega)$ is given by
the matrix element of the full Hamiltonian evaluated for the
propagating
state $|1\rangle\!=\!\sqrt{\frac{2}{L}}\sum_{n}\exp(i 2nk_b)
a_{2 n \hat{\bf b}}|{\rm AO}\rangle$, where $L$ is the system length
along $\hat\textbf{b}$, and
the hole is at the origin in Fig. \ref{Fig3}(a).

The full Green's function contains the selfenergy,
$G^{-1}(k_b,\omega)=G_0^{-1}(k_b,\omega)-\Sigma(\omega)$, which
stems from the confined motion initiated by the first step in the
$\textbf{a}$ direction transverse to the coherent propagation
along $\textbf{b}$. The confined motion [Figs. \ref{Fig3}(a)-(d)]
is accompanied by string formation and path retracement by holes.
We evaluated $\Sigma(\omega)$ analytically by applying the
recursion procedure, i.e. by the consecutive action with the
Hamiltonian (\ref{t2gmodel}) on the state $|1\rangle$, which
represents the hole created in the perfect AO order [Fig.
\ref{Fig3}(a)].
The only approximations consist of neglecting:
(i) all details in long hole paths,
(ii) some processes mediated
by the term (\ref{H3sl}) (such as those represented by Figs.
\ref{Fig3}(c) and \ref{Fig3}(f); they bring about only a minor
incoherent contribution \cite{Woh08}), and (iii) hole-hole
interactions within AO domains. On the other hand, we always
implement the constraint that the hole path can not encircle a
plaquette --- its topological reason can be recognized by
analyzing Fig. \ref{Fig3}.

\begin{figure}[t!]
 \includegraphics[width=8.4cm]{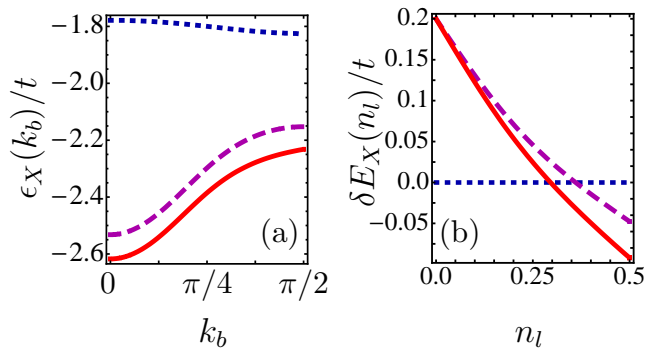}
\caption{(color online)
(a) Quasiparticle dispersion $\epsilon_X(k_b)$ for the
homogeneous AO phase ($X$=0, dotted) and for two kinds of stripes:
$X$=PW (dashed) and $X$=DW (solid line).
(b) Energy gain
$\delta E_X(n_l)$ per site (\ref{stripes}) for two stripe phases
for increasing stripe filling $n_l$. Parameters: $J=0.4 t$,
$\tau=0.1 t$. } \label{Fig4}
\end{figure}

$\Sigma(\omega)$ takes the form of a continued fraction.
For example, if we neglect, for demonstration purposes only the term
(\ref{H3sd}) (it is taken into account in the actual calculation but
gives only minor corrections as $\tau\ll t$),
one finds
\begin{equation}
\Sigma(\omega)=\frac{2 t^2}{\omega-\frac{7}{4}J
-\frac{2 t^2}{\omega-9J/4-\ldots}}\,,
\label{Sigma}
\end{equation}
where prefactors $2$ at $t^2$ refer to the number of ways by which
a path can be further extended after the 1st and 2nd hop to a
neighboring site. Such terms represent off-diagonal matrix
elements of the Hamiltonian between consecutive states created
during the recurrence procedure \cite{Sta94}. The prefactors $7$
and $9$ at $J/4$ stand for the number of bonds for which the
superexchange (\ref{HJ}) gives $0$ for paths of length 1 and 2
respectively (instead of $-J/4$ as for all bonds in the $|{\rm
AO}\rangle$ state)
--- they enter as diagonal matrix elements of the
Hamiltonian. By searching for zeros of $G^{-1}(k_b,\omega)$ the
quasiparticle dispersion $\epsilon_0(k_b)$ can be derived. For the
numerical results shown in Fig. \ref{Fig4}(a) sufficiently long
paths (up to 12 steps) have been considered in order to obtain the
saturation of results.

The same method of analysis apart from a minor modification may be
also applied to the DW stripe itself. Since two states shown in Figs.
\ref{Fig2}(b) and \ref{Fig2}(c) are not related to each other by a
translation, both of them,
$|1'\rangle\!=\!\sqrt{\frac{2}{L}}\sum_{n} \exp(i 2 n k_b) a_{2n
\hat{\bf b}} |{\rm DW}\rangle$ and
$|1''\rangle\!=\!\sqrt{\frac{2}{L}}\sum_{n} \exp\{i (2n+1) k_b\}
a_{(2n+1) \hat{\bf b}} |{\rm DW}\rangle$, need to be chosen as a
starting point in the recursion method. Here
$|{\rm DW}\rangle$ stands for the empty stripe depicted in Fig.
\ref{Fig2}(a) and the position of the hole in Fig. \ref{Fig2}(b)
is given by the zero vector.
Thus the recursion method has to be generalized to the case of
several initial states. We have found this generalization, but
since the procedure has turned out to be equivalent to the so
called projection-operator technique \cite{Mor65}, we do not
discuss it here in detail. The Green's function and the
self-energy are given now by $2\times 2$ matrices. The inverse
bare Green's function has only off-diagonal nonvanishing matrix
elements,
$[G_0^{-1}(k_b,\omega)]_{1'1''}=[G_0^{-1}(k_b,\omega)]_{1''1'}=
\omega-2t \cos(k_b)-J/4$,
where the reference energy due to the Ising term (\ref{HJ}) has been
subtracted again.
When applying the recurrence procedure to
the evaluation of ${\bf \Sigma}(\omega)$ we may use our previous
observations regarding the matrix elements of the Hamiltonian between
states representing different paths, in some cases including
modifications brought about by the presence of the DW.

The PW stripe formed as a FO ordered chain within the AO order, shown
in Fig. \ref{Fig2}(f), can be also solved by using the generalized
version of the recursion method. By looking for the zeros of the
inverted Green's function, or of the determinant, if $G(k_b,\omega)$
is given by a matrix, we have determined and showed in Fig.
\ref{Fig4}(a) the energy dispersion $\epsilon_0(k_b)$ in the
$\hat\textbf{b}$ direction for the homogeneous system with AO order,
$\epsilon_{\rm PW}(k_b)$ for the PW stripe, and
$\epsilon_{\rm DW}(k_b)$ for the antiphase DW stripe. Different
dispersions confirm that the hole motion stems from $t$ ($\tau$)
hopping in the stripe phases (homogeneous AO phase).

We investigate the stability of both types of stripe phases using
the energy gain per site ($X={\rm DW,PW}$),
\begin{equation}
\delta E_X(n_l)=\frac{1}{\pi}\int_0^{n_l\pi}d k_b\;\epsilon_X(k_b)
-E_0(n_l)+\delta E_J \,, \label{stripes}
\end{equation}
with respect to the doped homogeneous AO phase with energy
$E_0(n_l)=n_l\epsilon_0(\pi/2)$, given by the band energy minimum
$\epsilon_0(\pi/2)$ and proportional to the linear stripe filling
$n_l$ in the low doping regime (when the linear filling does not
increase the global filling). Here $\delta E_J=J/2$ is the energy
cost of two broken bonds in a stripe phase. The stripes are
stabilized by increasing $n_l$, see Fig. \ref{Fig4}(b), but the PW
stripes are somewhat less stable which we interpret as following
from the destructive interference of hole penetration paths into
left/right AO domains.

In conclusion, we have shown that a purely electronic mechanism leads
to self-organization in the form of FO ordered stripes at antiphase
DWs, penetrating into the AO order --- this novel phase becomes more
stable than the doped homogeneous AO state at the linear filling of
$n_l\simeq 0.26$ ($n_l\simeq 0.25$) for $J=0.4 t$ ($J=0.2 t$).
The energy gain over the PW stripes is small for all $n_l$, which
suggests that also the latter form of stripes might be formed at
finite temperature. These features are unique and can be of relevance
to the behavior of doped Mott insulators with $t_{2g}$ or $p$ orbital
order when spins may be neglected.

We thank P. Fulde, S. Kirchner, and K. Wohlfeld for insightful
discussions. A.M.O. was supported by the Foundation for Polish
Science (FNP) and by the Polish Ministry of Science and Education
Project N202 068 32/1481.



\begin{thebibliography}{99}
\bibitem{noteu} This condition is fulfilled in transition metal
                oxides; here
                an effective interaction $U$ includes Hund's exchange.
\end{thebibliography}
\end{document}